\documentclass[12pt,a4paper,english,preprint,aps,nofootinbib]{revtex4}
\usepackage[english]{babel}
\usepackage{graphicx,color}
\usepackage{latexsym}
\usepackage{amsmath}
\usepackage{amsfonts}
\usepackage{amssymb,slashed}
\usepackage{bbold}
\usepackage[latin1]{inputenc}
\usepackage{verbatim}

\DeclareMathOperator{\Tr}{Tr}

\begin{document}

\title{Electric Dipole Moment from QCD $\theta$ and How It Vanishes for Mixed States}

\author{\textbf{A. P. Balachandran}\footnote{bal@phy.syr.edu} $^{a,b}$}

\author{\textbf{T. R. Govindarajan}\footnote{trg@imsc.res.in} $^{b}$}

\author{\textbf{Amilcar R. de Queiroz}\footnote{amilcarq@unb.br} $^{b,c}$}  

\affiliation{$^a$ Physics Department, Syracuse University, Syracuse, NY, 13244-1130, USA}

\vspace*{1cm}

\affiliation{$^b$ Institute of Mathematical Sciences, CIT Campus, Taramani, Chennai
600113, India}

\vspace*{1cm}

\affiliation{$^c$ Instituto de Fisica, Universidade de
Brasilia, Caixa Postal 04455, 70919-970, Brasilia, DF, Brazil}

\preprint{IMSc/2012/4/8}

\begin{abstract}
In a previous paper \cite{Balachandran:2012fq}, we studied the $\eta'$ mass and formulated its chirally symmetric coupling to fermions which induces electric dipole moment (EDM). Here we calculate the EDM to one-loop. It is finite, having no ultraviolet divergence while its infrared divergence is canceled by soft photon emission processes \emph{exactly} as for $\theta=0$. The coupling does not lead to new divergences (not present for $\sin\theta=0$) in soft photon processes either. Furthermore, as it was argued previously \cite{Balachandran:2012fq}, the EDM vanishes if suitable mixed quantum states are used. This means that in a quantum theory based on such mixed states, a strong bound on EDM will not necessarily lead to a strong bound such as $|\sin \theta|\lesssim 10^{-11}$ . This fact eliminates the need to fine-tune $\theta$ or for the axion field.
\end{abstract}
\maketitle

\tableofcontents

\section{Introduction}

A non-zero electric dipole moment (EDM) of a nucleon implies parity ($P$-) and time-reversal ($T$-) violations. In the conventional approach to QCD, its $\theta$-term in the action, which violates $P$ and $T$, induces an electric dipole moment $d_N$ of the neutron. For small $\theta$, it is \cite{Veneziano:1979ec,Crewther:1979pi,E1980363,Dar2009,PospelovAnnalsPhys.318:119-1692005}
\begin{equation}
      |d_N|\approx \theta \cdot 2 \cdot 10^{-16}~\rm{cm}.
\end{equation}

On the other hand, the current experimental bound on $d_N$ is
\begin{equation}
      |d_N|< 6 \cdot 10^{-26}~\rm{cm}~(90 \%~\textrm{confidence level}),
\end{equation}
suggesting that $\theta$ is nearly zero:
      \begin{equation}
            |\theta|\lesssim 10^{-10}~\textrm{radians}.
      \end{equation}
The need for such fine-tuning of $\theta$ is the strong $CP$-problem.      

A point of this work is that $d_N$ vanishes for any $\sin\theta$ if appropriate mixed states are used. The general considerations of our previous work \cite{Balachandran2012} is thus confirmed by an explicit calculation.

As a first step to show this result, we also implement a chiral model calculation of EDM for the conventional QCD $\theta$-vacuum. It is finite, having no ultraviolet divergence while its infrared divergence is canceled by soft photon emission processes \emph{exactly} as for $\theta=0$. The coupling does not lead to new divergences (not present for $\sin\theta=0$) in soft photon processes either. We believe that this model which fully incorporates chiral symmetry merits attention and can be a useful tool to investigate $P$- and $T$-violations.

Our general one-loop result for the $P$- and $T$-violating fermion-photon vertex is reported in later sections. For photons on mass shell, and for the conventional pure QCD states, it gives the EDM (see equation (\ref{neutron-EDM-1}) ahead)
\begin{equation}
      |d_N|= 8\pi\alpha~\frac{\mu_2|\sin\theta|}{M^2}
\end{equation}
for a fermion of mass $M$, with $\alpha=e^2/4\pi=1/137$. It depends also on a new mass scale $\mu_2$.

In section II, we briefly recall our results from the earlier paper \cite{Balachandran:2012fq}. In particular, we explicitly show that the model contains the analogues of ``covariant'' and ``consistent'' axial charges \cite{Bertlmann:1996xk,Harvey2007}. As in QCD, while the former is not conserved, the latter is. 

The Dirac equation and the fermion propagator are modified by the $\theta$-angle. They are also described in section II. We also need an expression showing the dependence  of Dirac spinors on $\sin\theta$. We find that as well in section II. Finally in this section, we sketch perturbation theory and show that the $\theta$-angle modifies \emph{either} the Dirac propagators \emph{or} the external spinors, but \emph{not both}. 

Section III takes up the analysis of the one-loop diagram for EDM. It is ultraviolet finite. It contains an infrared divergent contribution which is canceled by soft photon emission processes \emph{exactly} as in standard QED. The remaining $\mathcal{O}(\alpha)$ contribution to EDM is evaluated to \emph{all} orders in $\cos\theta$.

In section IV, we recall the use of appropriate mixed states and how it alleviates $P$- and $T$-violations due to non-vanishing $\sin\theta$. It is then applied to the QCD $\theta$-angle to show that EDM vanishes for these mixed states.

In the final section V, we make general remarks in the use of mixed states for the restoration of anomalous symmetries.

\section{Resum\'e and Remarks on Previous Work}

In this section, we summarize the pertinent results from our previous work \cite{Balachandran:2012fq} on the $\eta'$ effective Lagrangian. We also point out how important features of the QCD axial anomaly are mirrored in this model.

In the Hamiltonian approach, the above $\eta'$ model has the following pairs of canonically conjugate fields:
      \begin{align}
	    \label{canonical-pairs-1}
            (\eta',  \pi) ~~~ \textrm{ and } ~~~  (B^\mu, P_\lambda).
      \end{align}
In addition, it has the first class constraints
\begin{align}
      \label{constraints-1}
      G(c) &= \int d^3x~(\partial_i c^i)(\vec{x})~\left(P_0-\lambda \eta' \right)(\vec{x},t), \\
      G'(w) &= \int d^3x~w^i(\vec{x})~P_i(\vec{x},t), \label{constraints-2}
\end{align}
where $c^i,w^i$ are Schwartz functions.

When there are $N_f$ flavors, chiral symmetry breaking implies the existence of an $N_f\times N_f$ matrix of fields 
\begin{equation}
      \label{fields-1}
      u=s~e^{i\eta'},
\end{equation}
where 
\begin{equation}
      \label{fields-2}
      u(x)\in U(N_f), ~~~ s(x) \in SU(N_f), ~~~ e^{i\eta'}(x)\in U(1).
\end{equation}
The field $s$ describes the $(N_f^2-1)$-dimensional flavor multiplet and $e^{i\eta'}$ describes the $\eta'$ field as in (\ref{canonical-pairs-1}).

The chiral group $SU(N_f)_L\times SU(N_f)_R$ acts on $u$ according to the rule
\begin{align}
      \label{chiral-transf-1}
      (g_L,g_R) \triangleright u(x) &\to g_L~u(x)~g_R^\dagger, \\
       (g_L,g_R) &\in  SU(N_f)_L\times SU(N_f)_R,
\end{align}
while the action of the axial vector group is
\begin{equation}
      \label{axial-transf-1}
      u(x) \to \omega~u(x), ~~~ |\omega|=1.
\end{equation}

Now, the fields (\ref{fields-2}) are invariant under
\begin{align}
      \label{ZN-transf-1}
      s(x) &\to s(x)~e^{i k \frac{2\pi}{N_f}}, \nonumber \\
      \eta'(x) &\to \eta'(x) - k \frac{2\pi}{N_f}, \\
      \textrm{with } &~k \in \left\{0,1,2,...,N_f-1 \right\}. \nonumber
\end{align}
That means in particular that the constraints and observables in field theory should be invariant under (\ref{ZN-transf-1}). The constraints (\ref{constraints-1}, \ref{constraints-2}) already are so.

Let $D^i$ be smooth functions on $\mathbb{R}^3$ fulfilling the condition
\begin{equation}
      \lim_{r\to\infty} \int d\Omega~r^2~\hat{x}_i D^i(\vec{x})=1.
\end{equation}
Then
\begin{equation}
      Q(D)=\int d^3x~(\partial_i D^i)(\vec{x})~\Big(P_0-\lambda\eta'\Big)(\vec{x},t)
\end{equation}
commutes with the constraints. However it is not a constraint as $D^i$ are not all Schwartz functions. But it fails to be invariant under (\ref{ZN-transf-1}). However
\begin{equation}
      W(x)=s(x)~e^{-\frac{i}{\lambda} Q(D)} := s(x) V
\end{equation}
is invariant under (\ref{ZN-transf-1}). So $W(x)$ is an observable.

We need to consider the following observables as well in what follows:
      \begin{itemize}
            \item[a)] $\int d^3 x~B^0(\vec{x},t)$ and its exponential form
            \begin{equation}
                  U(\theta)=e^{i\theta\lambda \int d^3x~B^0(\vec{x},t)}.
            \end{equation}
            \item[b)] $\pi-\lambda B^0$.
      \end{itemize}

Let
      \begin{equation}
            |\cdot\rangle_{\theta=0} := |\cdot\rangle_0
      \end{equation}
denote any QCD $\theta$-state vector for $\cos \theta = 1$ or $\theta=0 \mod 2\pi$ (Hereafter, whenever we write $\theta$, we mean $\theta \mod 2\pi$.). Then the corresponding state vector for any $\cos\theta$ is
\begin{equation}
      U(\theta)~|\cdot\rangle_0.
\end{equation}

We can write the fermion coupling to chiral fields either for quarks or for baryons.

Let us first focus on quarks. The relevant couplings, chirally invariant, but with $P$ and $T$ violations are the quark mass terms:
      \begin{equation}
	    \label{mass-term-1}
            \mathcal{L}_q=\mu_1 \left( \bar{q}_L~u~q_R+\bar{q}_R~u^\dagger~q_L\right) + \mu_2 \left( \bar{q}_L~W~q_R+\bar{q}_R~W^\dagger~q_L\right).
      \end{equation}
The second term is new for $\theta\neq 0$. But for $\theta=0$, $W=u$ and (\ref{mass-term-1}) becomes the standard quark mass term with mass $\mu_1+\mu_2$.

Let
\begin{equation}
      Q_A = \int d^3x~(q^\dagger_L q_L - q^\dagger_R q_R)
\end{equation}
be the axial charge of quarks:
      \begin{align}
            e^{i\alpha Q_A}~q_L~e^{-i\alpha Q_A} &= e^{-i\alpha}~q_L, \\
            e^{i\alpha Q_A}~q_R~e^{-i\alpha Q_A} &= e^{+i\alpha}~q_R.
      \end{align}
It is not a symmetry of (\ref{mass-term-1}). But just as in QCD we can have a ``consistent'' axial charge \cite{Bertlmann:1996xk,Harvey2007} by the addition to $Q_A$ the following term:
      \begin{equation}
            Q_\pi = -2 \int d^3x~\pi(\vec{x},t).
      \end{equation}
Then since
\begin{equation}
      e^{i\alpha Q_\pi}~e^{-i\eta'}~e^{-i\alpha Q_\pi} = e^{2i\alpha}~e^{-i\eta'},
\end{equation}
we have that
\begin{equation}
      Q_A+Q_\pi=\textrm{ constant of motion}.
\end{equation}

\subsection{Perturbation Theory}

The QCD state vector $|\cdot\rangle_0$ is one which does not lead to $P$- or $T$-violation. Its vacuum vector $|0\rangle_0$ is $P$- and $T$-invariant. If $|\cdot\rangle_0$ is a bipartite or multi-partite system with gluons, quarks and photons, neither the gluon nor the quark state vectors in $|\cdot\rangle_0$ lead to $P$- or $T$-violation.

In our model, the constituents in the multi-partite system are chiral fields and quarks. The dressing transformation $U(\theta)$ does not contain quark fields. We will do perturbation theory in the electromagnetic coupling constant $e$. In that case, \emph{the free external quark state vectors are not twisted by $U(\theta)$, only the chiral fields are}. This point is important in what follows.

In our calculation of EDM, we do not consider propagating chiral fields. It is enough thus to focus on the terms in the Hamiltonian involving the fermion fields. Their kinetic terms and coupling to electromagnetism have no chiral fields and are not affected by $U(\theta)$. The terms in the Hamiltonian sensitive to $U(\theta)$ are the quark mass terms:
      \begin{equation}
	    \label{mass-term-2}
            h^0_{q}= - \int d^3x~\left[ \mu_1 \left( \bar{q}_L~u~q_R+\bar{q}_R~u^\dagger~q_L\right) + \mu_2 \left( \bar{q}_L~W~q_R+\bar{q}_R~W^\dagger~q_L\right) \right]
      \end{equation}

We now replace $h^0_{q}$ by
\begin{equation}
      U(\theta)^{-1}~h^0_{q}~U(\theta) \equiv h^\theta_{q} = - \int d^3x~\left[ \bar{q}_L \left(\mu_1~u + e^{-i\theta}\mu_2~ W \right) q_R +~\rm{h.c.} \right]
\end{equation}
in order that henceforth we can work with untwisted $\theta=0$ state vectors $|\cdot\rangle_0$.

In perturbation theory in $e$, to zeroth order, $|\cdot\rangle_0$ splits as follows:
      \begin{equation}
            |\cdot\rangle_0=| \textrm{chiral particles}\rangle_0 \otimes |\textrm{quarks and photons} \rangle_0.
      \end{equation}

As we do not consider excitations of chiral particles, we can replace the first factor by $|0\rangle_0$, the vacuum of the chiral fields. In this vacuum, with the usual boundary conditions
\begin{equation}
      u(\vec{x},t)\to 1 ~~~ \textrm{ as } ~~ |\vec{x}|\to \infty 
\end{equation}
we have
\begin{equation}
      _0\langle 0 | u(\vec{x},t) | 0 \rangle_0=~_0\langle 0 | e^{i\eta'(\vec{x},t)} | 0 \rangle_0=1.
\end{equation}

In contrast to $(48)$ of \cite{Balachandran:2012fq} which is appropriate for $\theta$-states, we now have also $_0\langle 0 | P_0(x) | 0 \rangle_0=0$.

The expectation value of $h^\theta_{q}$ for the vector $|0\rangle_0$ gives the following effective quark mass term $H^\theta_{q}$ with $\theta$-effects included:
      \begin{align}
            H^\theta_{q}=~_0\langle 0 | h^\theta_{q} | 0 \rangle_0 &=- \int d^3x~\left[ \bar{q}_L \left(\mu_1  + e^{-i\theta} \mu_2  \right) q_R +~\rm{h.c.} \right] \nonumber \\
            &= - \int d^3x~ \bar{q} \left(\mu_1  + e^{i\gamma_5\theta} \mu_2  \right) q. 
      \end{align}

      Set
      \begin{equation}
	    \label{mass-term-3}
            \mu_1  + e^{i\gamma_5\theta} \mu_2  = \mu(\theta) + i \gamma_5 \mu'(\theta),
      \end{equation}
where
\begin{align}
      \mu(\theta) &= \mu_1 + \mu_2 \cos \theta, \\
      \mu'(\theta) &= \mu_2 \sin \theta. \label{mu-prime-def-1}
\end{align}
Then
\begin{equation}
      \mu_1  + e^{i\gamma_5\theta} \mu_2=|M(\theta)|~e^{i\gamma_5\lambda(\theta)},
\end{equation}
with
\begin{align}
      |M(\theta)|^2 &=\mu(\theta)^2+\mu'(\theta)^2, \\
      \cos \lambda(\theta) &=\frac{\mu(\theta)}{|M(\theta)|}, \\
      \sin \lambda(\theta) &=\frac{\mu'(\theta)}{|M(\theta)|}.   \label{sin-lambda-def-1}   
\end{align}

\emph{The significance of $|M(\theta)|$ is that it is the quark mass renormalized by $\theta$-angle}. This can be seen from the poles of the quark propagator for the mass term (\ref{mass-term-3}):
      \begin{equation}
            S_F(k) = \frac{-i\gamma\cdot k+\overline{M}(\theta)}{k^2+|M(\theta)|^2-i\epsilon},
      \end{equation}
where
\begin{equation}
      \overline{M}(\theta)=|M(\theta)|~e^{-i\gamma_5 \lambda(\theta)}.
\end{equation}

For the external spinors, it is thus natural to use the $P$- and $T$-invariant Dirac equation for mass $|M(\theta)|$:
      \begin{equation}
            \Big( i\gamma\cdot k + |M(\theta)|\Big)~ u(k)=0.
      \end{equation}
We will do so hereafter.

\section{The Loop Diagram}

\begin{figure}[!ht]
  \begin{center}
         \includegraphics[scale=0.6]{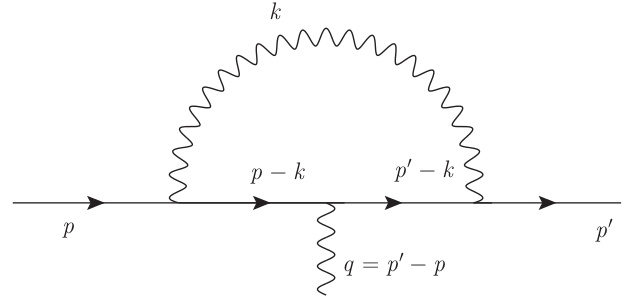}
   \end{center}
  \caption{Vertex correction diagram.} \label{fig-1}
\end{figure}

The diagram we must calculate is a simple modification of the one which corrects the electron magnetic moment. It is shown in figure \ref{fig-1}. Following Weinberg \cite{Weinberg:1995mt}, section 11.3, page 485, we obtain the vertex function
\begin{align}
      \label{vertex-1}
      \bar{u}(p')~\Gamma^\mu(p',p)~u(p)
\end{align}
where
\begin{align}
      \label{Gamma-0}
      \Gamma^\mu(p',p)=\int d^4 k &\left( e \gamma^\rho (2\pi)^4\right) \left[-\frac{i}{(2\pi)^4}\frac{-i\gamma\cdot (p'-k)+\overline{M}(\theta)}{(p'-k)^2+|M(\theta)|^2-i\epsilon} \right]\cdot  \\ &\cdot\gamma^\mu~\left[ -\frac{i}{(2\pi)^4}\frac{-i\gamma\cdot (p-k)+\overline{M}(\theta)}{(p-k)^2+|M(\theta)|^2-i\epsilon}\right] \left(e \gamma_\rho (2\pi)^4 \right)\left[-\frac{i}{(2\pi)^4}\frac{1}{k^2-i\epsilon} \right]. \nonumber
\end{align}

This differs in an essential way from Weinberg's equation (11.3.1) only in that $\overline{M}(\theta)$ has a $\gamma_5$-term. We can thus simplify $\Gamma^\mu(p',p)$ exactly as in that book arriving at the analogue of his (11.3.4):
      \begin{equation}
	    \label{Gamma-1}
            \Gamma^\mu(p',p)=\frac{2ie^2}{(2\pi)^4}\int_0^1 dx \int^x_0 dy \int d^4k \frac{\tilde{\Gamma}^\mu}{\left[k^2+|M(\theta)|^2 x^2+q^2y(x-y)-i\epsilon\right]^3}~, 
      \end{equation}
where $q=p'-p$ and
\begin{align}
      \label{Gamma-2}
      \tilde{\Gamma}^\mu = &\gamma^\rho \left[ -i\left(\gamma\cdot p'(1-y)-\gamma\cdot k - \gamma\cdot p (x-y) \right)+\overline{M}(\theta)\right] \nonumber \\  &\gamma^\mu~\left[-i\left( \gamma\cdot p(1-x+y)-\gamma\cdot k - \gamma\cdot p' y \right) + \overline{M}(\theta)\right] \gamma_\rho.
\end{align}

The $P$- and $T$-violating term comes from the dependence of $\overline{M}(\theta)$ in $\tilde{\Gamma}^\mu$ on $e^{-i\gamma_5 \lambda(\theta)}$. Only the homogeneous, first order terms in $\overline{M}(\theta)$ retain these factors: they cancel in the second order one because of the intervening $\gamma^\mu$:
      \begin{equation}
            \overline{M}(\theta) \gamma^\mu \overline{M}(\theta) = |M(\theta)|\gamma^\mu |M(\theta)|.
      \end{equation}
Retaining just first order $\overline{M}(\theta)$-terms in (\ref{Gamma-2}) and dropping $\gamma\cdot k$ terms (``symmetrical integration''), we obtain $\tilde{\Gamma}^\mu_{(1)}$:
      \begin{align}
	    \label{new-vertex-1}
            \tilde{\Gamma}^\mu \to \tilde{\Gamma}^\mu_{(1)} &= -i M(\theta)~\gamma^\rho \gamma^\mu \left[ \gamma\cdot p~(1-x+y)- \gamma\cdot p'~y\right] \gamma_\rho \nonumber \\
            &~-iM(\theta)~\gamma^\rho \left[ \gamma\cdot p'~(1-y) - \gamma\cdot p (x-y)\right] \gamma^\mu \gamma_\rho
      \end{align}

\subsection{Averaging over $x$ and $y$}

Following Weinberg \cite{Weinberg:1995mt}, suppose we change variables in the $x,y$ integrations in (\ref{Gamma-1}) according to
\begin{equation}
      x\mapsto x, ~~~~~~ y\mapsto x-y.
\end{equation}
This does not affect the denominator in (\ref{Gamma-1}), nor does it affect the $x$-, $y$-integration in that equation. If we now add to (\ref{Gamma-2}) its expression after these changes and divide by $1/2$, the linear terms in $\tilde{\Gamma}^\mu$ simplify considerably letting us replace linear $x$- and $y$-terms as follows:
      \begin{equation}
            x\mapsto x, ~~~~~~ y\mapsto \frac{1}{2} (y+x-y)=\frac{x}{2}.
      \end{equation}
Of course, this rule is not valid for quadratic expressions in $x$ and $y$, but by (\ref{new-vertex-1}), we do not have to deal with such terms.

Averaging over $x$ and $y$ simplifies $\tilde{\Gamma}^\mu_{(1)}$ to $\tilde{\Gamma}^\mu_{(2)}$:
      \begin{align}
            \tilde{\Gamma}^\mu_{(1)} \to \tilde{\Gamma}^\mu_{(2)}&= -i M(\theta) \gamma^\rho \gamma^\mu \left[ \gamma\cdot p - \gamma\cdot (p+p')\frac{x}{2}\right] \gamma_\rho \nonumber \\
            &~-iM(\theta) \gamma^\rho \left[ \gamma\cdot p' - \gamma\cdot (p+p') \frac{x}{2}\right] \gamma^\mu \gamma_\rho.
      \end{align}
Since $\gamma^\rho \gamma^\mu \gamma^\lambda \gamma_\rho=4 \eta^{\mu\lambda}$, we can simplify this further:
      \begin{equation}
            \tilde{\Gamma}^\mu_{(2)}=-8i~M(\theta)~(p+p')^\mu \left(1-\frac{x}{2}\right).
      \end{equation}

We now use the identity
\begin{equation}
      \int\frac{d^4 k}{(k^2+\Lambda)^3}=\frac{(2\pi)^4}{4\Lambda}
\end{equation}
to obtain
\begin{align}
      \label{current-correction-1}
      \bar{u}(p')~\tilde{\Gamma}^\mu_{(2)}~u(p) &= 4e^2~\bar{u}(p')~M(\theta) (p+p')^\mu~u(p) \nonumber \\ 
      &~\int_0^1 dx \int_0^x dy \frac{1-x/2}{|M(\theta)|^2x^2+ q^2 y(x-y)}.
\end{align}
The momentum dependence of this expression is governed by $(p+p')^\mu$ as required by current conservation.

Note that we have not encountered ultraviolet divergence in arriving at (\ref{current-correction-1}). But the constant term in the numerator of the integrand of this integral, that is, the term $1$,  leads to infrared divergence. It is \emph{exactly} the infrared divergence, which is homogeneous and linear in the electron mass, encountered in the anomalous magnetic moment calculation \cite{Schwinger:1948iu}. It is therefore \emph{exactly} canceled by the contribution from soft photon emission processes \cite{Weinberg:1995mt}. The twist $e^{i\gamma_5\lambda(\theta)}$ in $M(\theta)$ makes no difference.

We can see this in another way. The result (\ref{current-correction-1}) can be interpreted as coming from changing incident and outgoing wave functions to
\begin{align}
      \tilde{u}(p) &= e^{i\gamma_5 \frac{\lambda(\theta)}{2}}~u(p),\\
      \tilde{u}(p') &= e^{i\gamma_5 \frac{\lambda(\theta)}{2}}~u(p'),
\end{align}
and working with $|M(\theta)|$ and not with $M(\theta)$. In this way the twist $U(\theta)$ of $\eta'$- (or ``gluon-'') vacuum has been transferred to $u$'s. But this change does not affect the QED vertex: $\bar{u}(p') \gamma^\mu u(p) = \overline{\tilde{u}}(p') \gamma^\mu \tilde{u}(p)$. The cancellation of the infrared divergence in this transformed situation is explained in text books \cite{Weinberg:1995mt}. So it cancels here as well.

Dropping the constant term in (\ref{current-correction-1}), we obtain the finite answer
\begin{align}
      \label{current-correction-2}
      \bar{u}(p')~\tilde{\Gamma}^\mu_{(2)}~u(p)\to\bar{u}(p')~\tilde{\Gamma}^\mu_{(3)}~u(p)= -2e^2~\bar{u}(p')~|M(\theta)|e^{i\gamma_5\lambda(\theta)}~u(p)~(p+p')^\mu~F(q^2),
\end{align}
with
\begin{equation}
      \label{integral-qsquare-1}
      F(q^2)=\int_0^1 dx \int_0^x dy~\frac{x}{|M(\theta)|^2x^2+ q^2 y(x-y)}.
\end{equation}

Equation (\ref{current-correction-2}) has a $P$- and $T$-conserving part
\begin{equation}
      \bar{u}(p')~\tilde{\Gamma}^\mu_{(3),E}~u(p)=-2e^2~|M(\theta)| \cos\left(\lambda(\theta) \right)~\bar{u}(p')u(p)~(p+p')^\mu~F(q^2)
\end{equation}
which has $\mathcal{O}(\sin^2\theta)$ corrections to the standard electromagnetic vertex. It has also a $P$- and $T$-violating EDM term
\begin{equation}
      \bar{u}(p')~\tilde{\Gamma}^\mu_{(3),O}~u(p)=-2e^2~|M(\theta)| \sin\left(\lambda(\theta) \right)~\bar{u}(p')(i\gamma_5)u(p)~(p+p')^\mu~F(q^2).
\end{equation}

Since 
\begin{equation}
      F(0)=\frac{1}{|M(\theta)|^2},
\end{equation}
we obtain the static electric dipole moment
\begin{equation}
      \label{fermion-EDM-1}
      d= -2e^2~\frac{\sin\left(\lambda(\theta)\right)}{|M(\theta)|}.
\end{equation}

For the neutron, identifying $|M(\theta)|$ with the neutron mass $\mu_N$, we obtain its EDM $d_N$ as
\begin{equation}
      \label{neutron-EDM-1}
      d_N=-2e^2~\frac{\sin\left(\lambda(\theta)\right)}{\mu_N}=-8\pi\alpha~\frac{\mu_2\sin\theta}{\mu_N^2},
\end{equation}
where in the last equality we have used (\ref{sin-lambda-def-1}) and (\ref{mu-prime-def-1}).

If we recall that experimentally $|d_N| \lesssim 6\cdot 10^{-26}~\rm{cm}$, this gives the bound (for $\mu_N \approx 940~\rm{MeV}\approx 7.5\cdot 10^{12}~\rm{cm}^{-1}$, $\alpha\equiv e^2/4\pi \approx 1/137$)
\begin{equation}
      |\sin\left(\lambda(\theta)\right)|=\frac{\mu_N~|d_N|}{8\pi\alpha}\lesssim 2.5 \cdot 10^{-12} .
\end{equation}

\section{On Use of Mixed States}

We have recalled in Section II that if $|\cdot\rangle_{\theta=0}$ is any hadronic vector state for $\theta=0$, then the corresponding vector state for finite $\theta$ is
\begin{equation}
      |\cdot\rangle_{\theta}= U(\theta) |\cdot \rangle_{\theta=0}.
\end{equation}

The expression (\ref{neutron-EDM-1}) for the EDM comes from the $P$- and $T$-violating part of the matrix element
\begin{equation}
      \label{exp-value-current-1}
      ~_\theta\langle p | \overline{\psi} \gamma^\mu \psi | p \rangle_\theta.
\end{equation}
Here we have suppressed labels like spin components and charge.

The expression (\ref{exp-value-current-1}) is the expectation value of the current $\overline{\psi}\gamma^\mu \psi$ for the density matrix
\begin{align}
      \rho_\theta &= |p\rangle_\theta~_\theta\langle p|~~: \label{density-matrix-1}\\
      ~_\theta\langle p | \overline{\psi} \gamma^\mu \psi | p \rangle_\theta &= \Tr \rho_\theta \overline{\psi}\gamma^\mu \psi	    \label{exp-value-current-2}
\end{align}
(The spin sum on the RHS will introduce overall factors. We can ignore them.).

The $P$- and $T$-transform of $U(\theta)$ is $U(-\theta)$. Hence that of $\rho_\theta$ is $\rho_{-\theta}$.

Our proposal in a previous work \cite{Balachandran:2012fq} was to use the mixed state
\begin{equation}
      \hat{\rho}=\frac{\rho_\theta+\rho_{-\theta}}{2}
\end{equation}
as a possible mechanism to obtain zero EDM and solve the strong $CP$ problem.

We can now see that the proposal works. Since the matrix element (\ref{density-matrix-1}) is proportional to $\sin\theta$, it is clear that the EDM term in $\Tr \hat{\rho}~\overline{\psi}\gamma^\mu \psi$ is zero. That is so no matter what $\sin\theta$ is. This means that \emph{with the use of the above mixed state, there is no way to constraint $\sin\theta$ from EDM. There is no need for an axion either}. 

Note that 
\begin{equation}
      \rho_{-\theta}\neq P\rho_\theta P^\dagger = P \rho_\theta P.
\end{equation}
That is because in $\rho_{-\theta}$, we reverse only $\theta$, whereas $P$ will also change variables like momentum and helicity. For example,
\begin{equation}
      P|p\rangle_{\theta}~_\theta\langle p| P =|p_0,-\vec{p}\rangle_{-\theta}~_{-\theta}\langle p_0,\vec{p}|,
\end{equation}
whereas
\begin{equation}
      \rho_{-\theta}=|p\rangle_{-\theta}~_{-\theta}\langle p|.
\end{equation}

\subsection*{Remarks}

Since $|p\rangle_\theta$ is not normalizable, we should really work with
\begin{equation}
      \int \frac{d^3 k}{2|k_0|}~f(k,p)~|k\rangle_\theta = |f\rangle_\theta,
\end{equation}
where $f$ is a function with compact support in $k$ centered at $p$ and normalized to $1$:
      \begin{equation}
            \int \frac{d^3 k}{2|k_0|}~|f(k,p)|^2=1.
      \end{equation}

Let 
\begin{equation}
      \hat{\rho}_\theta(f)=|f\rangle_\theta~_\theta\langle f|.
\end{equation}
Then the mixed density matrix is
\begin{equation}
      \hat{\rho}(f)=\frac{\hat{\rho}_\theta(f)+\hat{\rho}_{-\theta}(f)}{2}.
\end{equation}
The EDM vanishes for $\hat{\rho}(f)$ as well:
      \begin{equation}
      \textrm{Terms odd under P and T in } \Tr \hat{\rho}_\theta(f)~\overline{\psi} \gamma^\mu \psi =0.
      \end{equation}
Hence our previous conclusions are not affected.

\section{Final Remarks}

There are good reasons to develop a quantum theory where certain symmetries do not become anomalous. Thus, consider the case dealt with in the present paper, namely parity and time-reversal anomaly in QCD when $\cos\theta\neq 1$. There is no experimental basis for this anomaly. It has led to the suggestion that there is a particle called the axion. There is no evidence for the latter either.

There are more of these examples. The following deserves special mention. The color group becomes anomalous in the presence of nonabelian magnetic monopoles \cite{Balachandran1984,Balachandran1984b,Balachandran2012}. That too is the fate of mapping class groups (``large diffeos") in certain quantum gravity theories \cite{Balachandran:2011gj}.

We may not always tolerate such anomalies. The use of mixed states is one way to recover them as symmetries.

The entropy created by these mixed states is not always small. The entropy of $\hat{\rho}(f)$ is roughly $\log 2$ and that is \emph{per} an elementary quantum state. It can easily be of the order of  $5$ or $10$ per quantum state depending on the volume of the anomalous group. In QCD plasma, entropy per particle is about 10 \cite{Wong:1995jf}. The mixed state entropy can be comparable to 10. They can thus affect phenomena in elementary particle physics.

Let $\sum_n \varphi_n \otimes \varphi_n^\dagger$ be the resolution of identity corresponding to the eigenstates of the Hamiltonian for a particular domain $\mathcal{D}$. If a symmetry such as parity is anomalous, we should average this resolution over the anomalous group. For parity in our problem, the averaged resolution of identity is
\begin{equation}
      \label{id-resolution-1}
      \frac{1}{2}\sum_n \left( \varphi_n \otimes \varphi_n^\dagger + P\varphi_n \otimes \varphi_n^\dagger P^{-1}\right).
\end{equation}
Call such an averaged resolution of identity in general as $\hat{\rho}$.

The use of anomaly-free mixed states implies that the Gibbs state is changed, the expectation values of correlators being 
\begin{equation}
      \langle \mathcal{O}_1 \cdots \mathcal{O}_N \rangle = \frac{\Tr \hat{\rho}~e^{-\beta H}~ \mathcal{O}_1 \cdots \mathcal{O}_N}{\Tr \hat{\rho}~e^{-\beta H}}.
\end{equation}
The different terms obtained by averaging in $\rho$ can affect these expectation values as they come from different domains of the Hamiltonian $H$.

This formula can also be adapted to thermofield theory \cite{Khanna:2009zz}. The properties of those correlators seem different from these for the standard Gibbs state.

\emph{Nonabelian} gauge symmetries $G$ which define ``charges" (as contrasted to those implied by Gauss law which hence vanish on vector states) have unusual properties \cite{balachandran1991classical}. Observables must commute with these symmetries. That means that only the elements in the center $\mathcal{C}(\mathbb{C} G)$ of the group algebra $\mathbb{C}G$ are observable.  By the preceding remark, if a gauge transformation $g\in G$ does not commute with  $\mathcal{C}(\mathbb{C} G)$, it is not observable. Restricting any pure state with nontrivial response to $G$ to its center $\mathcal{C}(\mathbb{C} G)$ creates a mixed state and entropy. We will develop this remark elsewhere. It already happens for the molecule $C_2H_4$ \cite{Balachandran1992c,balachandran1991classical}.

\section{Acknowledgement}

The authors would like to thank Alvaro Ferraz for the hospitality at the International Institute of Physics at the Universidade Federal do Rio Grande do Norte in Natal, Brazil where discussions that led to this work were initiated. APB and ARQ acknowledge the warm hospitality of Prof. Alberto Ibort at Departamento de Matem\'aticas, Universidad Carlos III de Madrid, Spain where part of this work was done. We thank Prof. S. Digal for discussion regarding QGP entropy. We also thank Prof. G. Marmo, Prof. K. Gupta, Prof. A. Reyes, Prof. S. Kurkcuoglu and Prof. X. Martin for discussions at the final stages of this work. APB is supported by DOE under grant number DE-FG02-85ER40231 and by the Institute of Mathematical Sciences, Chennai. ARQ is supported by CNPq under process number 307760/2009-0.


  
 \providecommand{\href}[2]{#2}\begingroup\raggedright\endgroup

\end{document}